\documentclass{jps-cp}
\usepackage{txfonts} 

\usepackage{array}

\usepackage{graphics}
\usepackage{graphicx}

\usepackage{color}
\definecolor{!R}{rgb}{1,0,0}
\definecolor{!G}{rgb}{0,1,0}
\definecolor{!B}{rgb}{0,0,1}
\definecolor{!W}{rgb}{1,1,1}

\newcommand{\ve}[1]{\mathbf{#1}}
\newcommand{\avr}[1]{\langle #1\rangle}
\newcommand{\intlim}[3]{\int\limits_{#2}^{#3}\!\!\mathrm{d}#1}

\usepackage{slashed}
\usepackage{empheq} 

\title{Non-static Analysis of the Anomalous Chiral Conductivities}

\author{Mikl\'os \textsc{Horv\'ath}$^{1}$, Defu \textsc{Hou}$^{1}$ and Hai-cang \textsc{Ren}$^{2,1}$}

\inst{$^{1}$Institute of Particle Physics and Key Laboratory of Quark and Lepton Physics (MOE), Central China Normal University, Wuhan 430079, China \\
$^{2}$Physics Department, The Rockefeller University, 1230 York Avenue, New York, New York 10021-6399, USA}

\email{miklos.horvath@mail.ccnu.edu.cn}

\abst{Given the intrinsic nonequilibrium nature of high-energy collisions the investigation of the dynamical properties of transport phenomena is important. The study of the real-time behavior of various conductivities and susceptibilities help refine the simulation tools we use to compare the theories and the experimental findings. In this contribution we take steps to give the chiral magnetic conductivity in case of magnetic field \textit{and} chiral imbalance are both space-time dependent. Using linear response approximation we present the general 1-loop resummed expression for the electric current. We also suggest simple limiting cases in hope for possible implementation into a hydrodynamical framework.}

\kword{chiral anomaly, CME, anomalous transport in the QGP}

\begin{document}
\maketitle

\section{Introduction}
The possible violation of parity in high-energy nuclear collisions (HEC) has been gaining an increasing attention in recent years. The large electromagnetic (EM) fields and huge vorticity present at the initial stage of the collision combined with the chiral imbalance of the QCD plasma at high temperatures might lead to anomalous transport phenomena such as the chiral magnetic (CME) and chiral vortical effects (CVE) -- see \cite{anomTransportRev,LandsteinerRev} for a review. \\
The CME is a transport phenomenon of genuinely quantum physical origin: the presence of electric current parallel to magnetic field in matter with chiral imbalance. It is crucial to have realistic simulations in hand in order to evaluate possible experimental signals of CME and CVE, as direct experimental evidence to local $\mathcal{CP}$-violation \cite{FukushimaLocalCP}. 
The chiral charge fluctuation of the QCD vacuum requires time and space dependent description of the transport coefficients. In this contribution we express the CME conductivity for arbitrary, space-time dependent external fields in terms of the fermionic spectral density. This allows us to calculate the medium response induced by even fast changing perturbations both in the chiral charge and in the magnetic field.\\
Hydrodynamics is proven most useful in modeling the early stages of the time evolution after the collision. Now this framework is being extended to take into account the chirality of the medium as well \cite{SonSorowka,AVFD1}. Without simulations, taking into account as many aspects of the HEC as possible, it is almost hopeless to one day understand how all these contributions come together in the actual observed signal. For this purpose	 it is important to extract the needed transport coefficients of the chiral matter and to understand its relaxation effects.

\subsection{Sources \& observables} 
In a non-central HEC, the participant nuclei induce EM fields. The magnetic field $\ve{B}$ is typically perpendicular to the reaction plane. Its lifetime is, however, highly uncertain. The quark-gluon plasma (QGP) with its chiral quark constituents is affected by the EM fields, so the resulting CME current could contribute to the formation of a nonzero dipole moment of the plasma. As it was shown \cite{polSTAR1,polSTAR2,polSTAR3,AVFD1}, the angular correlations of charged particles with either same or opposite electric charge are sensitive to the CME. Because of the inaccessibility of $\mathcal{CP}$-odd sensitive observables, one needs to measure two- and three-particle correlations on an event-by-event basis in order to obtain any CME signal. These correlators are $\mathcal{CP}$-even, therefore sensitive for example to the flow resulted by the hydrodynamic expansion of the medium. The subtraction of these accompanying background effects need careful treatment \cite{polSTAR1,polSTAR2,polSTAR3}.\\
Since the electric and chiral currents are coupled to each other, the anomaly also manifests itself by the plasma developing a quadrupole moment via the so called chiral magnetic wave (CMW) \cite{CMW1,CMW2}. Such waves might get attenuated near to a critical point \cite{isospinCMW}, offering another way of experimental detection of chiral restoration. The chiral imbalance of QGP originates from two sources: $i.)$ the EM-sector produces chiral charge with a rate proportional to $\ve{E}\cdot\ve{B}$, but $ii.)$ the anomaly affects the color gauge fields as well. It is theorized that, in the QGP, bubbles with different chiral charge are created out of the QCD vacuum, making the chiral charge density highly inhomogeneous. Charged partons traveling between plasma regions of different chirality can produce transition radiation. Yet another sign of chiral anomaly, circularly polarized photons are expected at a given angle with respect to the direction of a jet initiated by fast partons \cite{transRadSign}.

\section{CME in linear response}
In a chiral medium the electric and chiral currents, $\ve{J}$ and $\ve{J}_5$, respectively, have components parallel to the magnetic field $\ve{B}$ naturally. When both the magnetic field and the chemical potentials are constant -- or the system is in a steady state --, the corresponding conductivities are fully determined by the chiral anomaly:
\begin{align}
\ve{J} = & \sigma_\text{ohmic}\ve{E} + \mu_5 C_A \ve{B}, \label{CMEcurr}\\
\ve{J}_5 = & \sigma_\text{CESE}\ve{E} + \mu C_A \ve{B}. \label{CSEcurr}
\end{align}
The CME conductivity is in this case $\sigma_\text{CME}=\mu_5 C_A$. $C_A$ is defined by the the anomalous divergence $\partial_\mu J^\mu_5 =C_A\ve{E}\cdot\ve{B}$. It depends only on the charge and the number of fermion species, for Dirac-fermions: $C_A=\frac{e^2}{2\pi^2}$. The conductivities $\sigma_\text{ohmic}$ and $\sigma_\text{CESE}$ are not universal, rather determined by the interactions.\\
Here we focus on the CME conductivity, which encodes the relationship between the electric current $\ve{J}$, the magnetic field $\ve{B}$ and the chemical potential associated to the chiral charge, $\mu_5$. Realistically both $\mu_5$ and $\ve{B}$ are space-time dependent. 
The initial chiral charge comes from QCD vacuum fluctuations, which is inherently inhomogeneous -- since the color fields are such. 
Thus, the response is no longer given by Eqs. (\ref{CMEcurr}, \ref{CSEcurr}). The relaxation back to the steady currents is controlled by the interactions. 
Several attempts have already been made to explore the relaxation behavior in non-static $\ve{B}$ \cite{nsCMEKharzeev,nsCMEKharzeev2,HouRen2011,HouRen2017,CMErelaxFLTfermions,secondOrderTransResummed, nonLinRespKinTh1,nonLinRespKinTh2,WignerFunc1,WignerFunc2,WignerFunc3}. Although, there have been studies investigating the possible inhomogeneity and time dependence of $\mu_5$ in terms of kinetic theory for example \cite{earlyMu5Kin1,earlyMu5Kin2}, this aspect is still not addressed thoroughly.\\
The electric and chiral currents $\avr{J^\mu_{(5)}}=\avr{\overline{\psi}\gamma^\mu(\gamma^5)\psi}$ are given in the linear response approximation. This way it is possible to keep all external fields space-time dependent, namely the EM vector potential $A_\mu(x)$ and the (axial) chemical potential $\mu_{(5)}(x)$. The structure of the currents are schematically the following:
\begin{align}
\avr{J^\mu} \sim & \avr{J^\mu J^\nu J_5^0}A_\nu\mu_5\,\, \left(\,\,+ \avr{J^\mu J^0}\mu + \avr{J^\mu J^\nu}A_\nu + \avr{J^\mu J^0 J_5^0}\mu \mu_5 \right) \label{structJlinres}\\
\avr{J_5^\mu} \sim & \avr{J_5^\mu J_5^0}\mu_5 + \avr{J_5^\mu J_5^0 J_5^0}\mu_5\mu_5 \,\,\left(\,\,+ \avr{J_5^\mu J^\nu J^\rho}A_\nu A_\rho + \avr{J_5^\mu J^0 J^0}\mu\mu +\avr{J_5^\mu J^0 J^\nu}\mu A_\nu\right) \label{structJ5linres}
\end{align}
Naturally, the external fields are needed to be convolved with the correlation function in the above expansion. All the averages are meant as the quantum system in question is at finite temperature, but all external fields are set to zero. 
The terms in the parenthesis contribute only for $\ve{E},\,\mu\neq 0$. If inhomogeneous, $\mu_5$ plays an important role even for zero $\ve{E}$ and $\mu$ (neutral plasma), as there are gradient corrections -- both time and space -- in $\mu_5$  in Eq. (\ref{structJ5linres}). The CME electric current in neutral plasma is given by
\[\avr{J^\mu}(x) =-\int_{q_1}\int_{q_2}\widetilde{A}^\text{ext}_\nu(q_1)\widetilde{\mu}_5(q_2)
i\delta G^{0\mu\nu}_{AVV}(q_1,q_2)e^{i x\cdot(q_1+q_2)}. \label{Jlinres}\]
This expression simply becomes a product, defining the static point: $i\delta G^{0ij}(q_1\rightarrow 0, q_2=0) =i\epsilon^{0ijk}q_{1k}C_A$ and leading to Eq. (\ref{CMEcurr}) when both external fields are set to zero.

\section{Non-static AVV triangle}
For the response currents in Eqs. (\ref{structJlinres}, \ref{structJ5linres}) one needs the two-point functions $\avr{JJ}$, $\avr{J_5J_5}$ and the three-point functions $\avr{JJJ_5}$, $\avr{J_5J_5J_5}$. This is equivalent to computing 1-loop corrections, which also can be resummed using the full fermion propagator. 
Now we focus on the AVV vertex function -- for $\ve{E}=0$ there is no contribution from $\avr{JJ}$ to the CME conductivity anyway. Using the Schwinger--Keldysh formalism, the Fourier transform of the AVV vertex function is given at one-loop order as follows:
\begin{align}
i\delta G_{AVV}^{\rho\mu\nu}(q_1,q_2) =
-\frac{ie^2}{2}\int_p\text{tr}& \left\{\gamma^\mu iG^C(p+q_1+q_2)\gamma^\rho\gamma^5 iG^A(p+q_1)\gamma^\nu iG^A(p) +\right. \nonumber\\
+& \gamma^\mu iG^R(p+q_1+q_2)\gamma^\rho\gamma^5 iG^C(p+q_1)\gamma^\nu iG^A(p) + \nonumber \\
+& \gamma^\mu iG^R(p+q_1+q_2)\gamma^\rho\gamma^5 iG^R(p+q_1)\gamma^\nu iG^A(p) + \nonumber \\
+& \left.\left(q_1\leftrightarrow q_2,\,\, \gamma^\rho\gamma^5\leftrightarrow\gamma^\nu\right) \right\}, \label{AVV1loop}
\end{align}
where $G^{R,A,C}$ are the retarded, advanced propagators and the correlator, respectively: $iG^{R/A}(x)=\pm\theta(\pm x_0)\rho(x)$, $iG^C=\avr{\{\overline{\psi},\psi\}}$, with the fermionic spectral function being $\rho=\avr{[\overline{\psi},\psi]}$. In the spirit of the linear response approximation, we evaluate the averages in thermal equilibrium, with external fields are set to zero. All propagators are linked through the spectral function in this case: $iG^C(p)=(1-2n_{FD}(p_0/T))\rho(p)$. The AVV vertex therefore can be expressed solely in terms of $\rho$:
\begin{align}
i\delta G_{AVV}^{\rho\mu\nu}(q_1,q_2) =&\frac{ie^2}{4\pi^2}\intlim{\Omega_1}{-\infty}{\infty}
\intlim{\Omega_2}{-\infty}{\infty}\int_p N(p_0/T,q_{10},q_{20},\Omega_1,\Omega_2) \nonumber \\
\times & \left(\text{tr}\left\{\gamma^\mu\rho(\Omega_1+p_0,\ve{p}+\ve{q}_1+\ve{q}_2)
\gamma^\rho\gamma^5\rho(\Omega_2+p_0,\ve{p}+\ve{q}_1)\gamma^\nu
\rho(p_0,\ve{p})\right\} -\left\{\text{same, with mass }M\gg q_1,q_2\right\}\right) \label{AVVrho}
\end{align}
Above the thermal function $N$ carries the effects of retardation, $n_{FD}$ being the Fermi-Dirac function:
\begin{align}
N = 2\frac{n_{FD}((p_0+\Omega_1)/T)(q_{10}-\Omega_2)+n_{FD}((p_0+\Omega_2)/T)(\Omega_1-q_{10}-q_{20})+n_{FD}(p_0/T)(q_{20}-\Omega_1+\Omega_2)}{(\Omega_1-q_{10}-q_{20}-i0^+)(\Omega_1-\Omega_2-q_{20}-i0^+)(\Omega_2-q_{10}-i0^+)}. \label{AVVthermfac}
\end{align}
Also it is important to mention that Eq. (\ref{AVVrho}) contains a Pauli--Villars (PV) contributions, since regularization is needed to ensure that Eq. (\ref{AVV1loop}) fulfills the Ward--Takahasi identities. 
See \cite{HouRen2017} from the point of view of the Wigner function, which also leads to Eq. (\ref{AVV1loop}) and uses PV regulators.

\section{CME conductivity in the homogeneous limit}
\begin{figure}[!h]
	\begin{tabular}{m{0.45\linewidth}m{0.45\linewidth}}
		\includegraphics[width=1\linewidth]{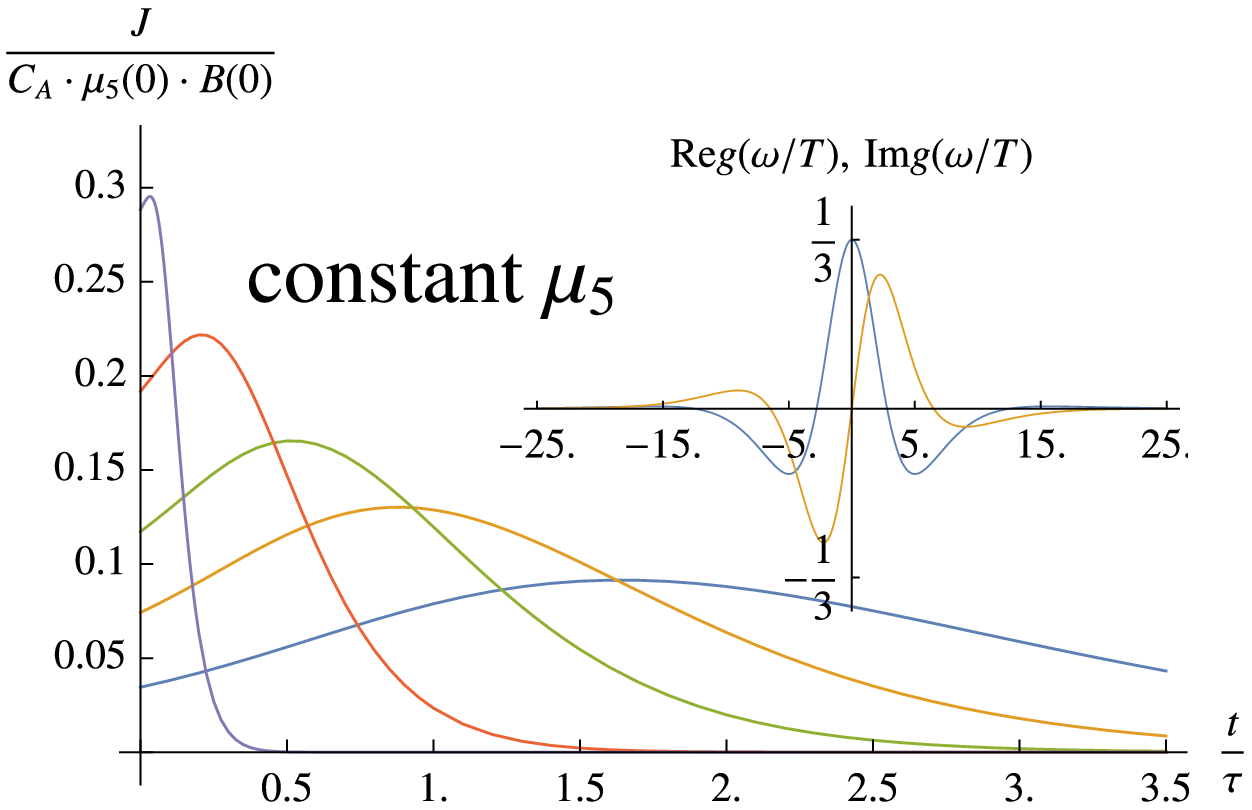} &
		\includegraphics[width=1\linewidth]{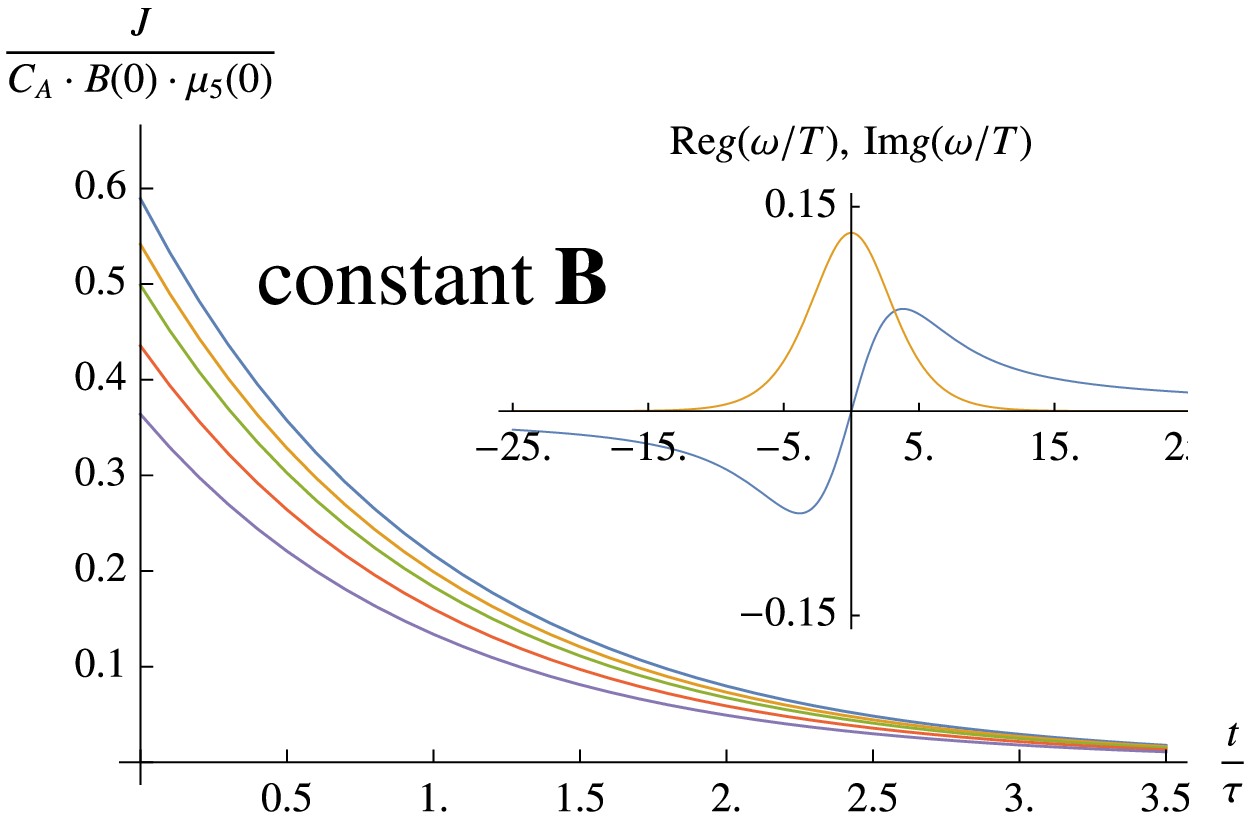}
	\end{tabular}
	\caption{(left) static $\mu_5$, homogeneous $B$; (right) static $\ve{B}$, homogeneous $\mu_5$ -- first $\ve{q}_1+\ve{q}_2=0$ was set, then $\ve{q}_1\rightarrow 0$. The colored curves refer to the same temperature values on both figures, $\tau T=0.2, 0.33, 0.5, 1.0, 4.0$ for blue, orange, green, red, purple, respectively. On the inset the real (imaginary) part is drawn by blue (yellow).}\label{fig:homCurrent}
\end{figure}

Here we give an example how the non-static behavior of the CME conductivity plays itself out in specific cases -- using noninteracting, chiral fermions. In Fig. \ref{fig:homCurrent} can be seen the resulting $\ve{J}$ of homogeneous fields, for $i.)$ static $\mu_5$ on the left and $ii.)$ static $\ve{B}$ on the right. In both cases, we assumed suddenly switched on $\ve{B}$ ($\mu_5$) at $t=0$, which decays exponentially with lifetime $\tau$: $\left\{B(t),\mu_5(t)\right\}= \left\{B(0),\mu_5(0)\right\}\theta(t)e^{-t/\tau}$. Case $i.)$ shows sizable delay in response, suppressed more as temperature decreases -- in agreement with \cite{nsCMEKharzeev}. On the contrary, case $ii.)$ shows no delay in response to the decay of $\mu_5$, and this response is suppressed as $T$ \textit{rises}. Also, the maximal value of the current is not larger than $1/3$ and $2/3$ of the static value in cases $i.)$ and $ii.)$, respectively. This also reflects the peculiarity of the different orders of approaching the static limit. While we send $\ve{q}_1$ to zero first and set $q_{10}=0$ in case $i.)$, in case $ii.)$ first the scale of inhomogeneity of $\ve{B}$ and $\mu_5$ was set at the same, invoking $\ve{q}_1+\ve{q}_2=0$, then $\ve{q}_1$ was sent to zero and finally $q_{10}=0$. This unambiguity of the homogeneous limit deserves a comprehensive analysis which we will give in details elsewhere \cite{HouHorRen2019}. \\
The frequency spectra $g$ of the conductivity, defined as $i\delta G^{0ij}_{AVV}(\ve{q}_1\rightarrow 0,\ve{q}_2=0,q_{10},q_{20})=\epsilon^{ijk}q_{1k} C_A(g_\text{stat.}+g(q_{10},q_{20}))$ behaves rather differently in the mentioned two limits, as showed in the insets of Fig. \ref{fig:homCurrent}. While its real (imaginary) part is even (odd) in frequency in case $i.)$, it is the opposite in $ii.)$ There is an additional static contribution to $g$, which in case $i.)$ does not matter, in case $ii.)$ it is $g_\text{stat.}(\omega)=\frac{1}{3}\frac{\omega}{\omega+i0^+}$ it adds an extra $\frac{1}{3}C_A B \mu_5(0) e^{-t/\tau}$ contribution to the response current.

\section{Conclusions and perspectives}
We gave expression for the general CME conductivity in terms of the fermion spectral function in Eq. (\ref{AVVrho}). We showed examples for non-static CME responses in two limiting cases -- static $\mu_5$ ($\ve{B}$), decaying $\ve{B}$ ($\mu_5$), respectively. Our treatment is also able to incorporate interactions between the fermions. This direction deserves further investigation. By introducing finite lifetime to the fermions the sensitivity to the different orders of limits, as approaching the static point, can be resolved to a smooth crossover, as suggested in \cite{CMErelaxFLTfermions}. \\
Other possible use of our results can be in hydrodynamic simulations. Firstly, the nonlocal convolutions of the conductivity and the external fields needed to be mapped into a local gradient expansion -- as usually relaxation dynamics is introduced into hydrodynamics. This can be done by high-$T$ expansion, which renders the Fourier transformed conductivity to the small momentum region. \\
In perspective, we plan to investigate the CME in case of inhomogeneous chiral charge $n_5$. For that, we need to model the generation of $n_5$ realistically based on the properties of QCD, using stochastic sources for example. This way it is possible to explore how the inhomogeneity of $n_5$ is translated into final state angular correlations via the full hydrodynamic evolution. \\
\textit{Acknowledgements.} This research was in part supported by the Ministry of Science and Technology of China (MTSC) under the ``973'' Project No. 2015CB856904(4) and by the NSFC under Grant Nos. 11735007, 11890711, 11847242.

\end{document}